\documentclass[11pt,a4paper]{article}
\usepackage[english]{babel}
\usepackage{verbatim}
\usepackage{graphicx}
\usepackage{amsfonts,epsfig}
\usepackage{amsmath}
\date{}

\begin{document}
\vspace*{0cm}
\begin{center}
{\setlength{\baselineskip}{1.0cm}{ {\Large{\bf DARBOUX TRANSFORMATIONS FOR \\ A GENERALIZED DIRAC EQUATION \\ IN
TWO DIMENSIONS \\}} }}
\vspace*{1.0cm}
{\large{\sc{Ekaterina Pozdeeva$^\dagger$}}} and {\large{\sc{Axel Schulze-Halberg$^\ddagger$}}}
\end{center}
\indent \\ $\dagger~$ICP RAS, Kosygina str. 4, 119991, Moscow, Russia, Email: ekatpozdeeva@mail.ru, pozdeeva@chph.ras.ru
\noindent \\ \\
$\ddagger~$Department of Mathematics and Actuarial Science, Indiana University Northwest, 3400 Broadway,
Gary IN 46408, USA, e-mail: xbataxel@gmail.com

\vspace*{1cm}

\selectlanguage{english}
\begin{abstract} \noindent
We construct explicit Darboux transformations for a generalized, two-dimensional
Dirac equation. Our results contain former findings for the one-dimensional, stationary Dirac equation, as well as for the fully
time-dependent case in (1+1) dimensions. We show that our Darboux transformations are applicable to the two-dimensional
Dirac equation in cylindrical coordinates and give several examples.

\end{abstract}
\noindent \vspace{.1cm}
\section{Introduction}
The famous Darboux transformation \cite{D1} \cite{D2} is a mathematical
scheme that interrelates solutions of differential equations by means
of a linear differential operator.  Since the Darboux transformation
does not involve changes of variables, it is essentially different from
other solution-generating methods, such as approaches related to Lie
symmetries. While the original Darboux transformation applied to
linear, one-dimensional equations of second order only, in the meantime
it has been generalized to a variety of linear and nonlinear models in
higher dimensions \cite{M} \cite{darbbook}. While such nonlinear
equations include the nonlinear Schr\"odinger equation, the nonlocal
Korteweg-de-Vries equation or the sine-Gordon equation, in the context
of linear models Schr\"odinger-type equations and the Dirac equation
are particularly important. The Darboux transformation has been
extensively applied to Schr\"odinger equations, especially since it has
been found to be equivalent to the quantum-mechanical supersymmety
formalism, see \cite{cooper} and references therein. As far as the
Dirac equation is concerned, a Darboux transformation has been
constructed for both the one-dimensional stationary case
\cite{samsonov} and the fully time-dependent equation \cite{samsonov2}
\cite{Yurov} in (1+1) dimensions. A further generalization of the
Darboux transformation to Dirac equations in higher spatial dimensions
will be made in the present work. In particular, we focus on a
generalized Dirac equation in two variables, a special case of which is
the stationary, two-dimensional Dirac equation. The Darboux
transformation for our generalized equation will be constructed by
means of an approproate intertwining relation (section 2), such that
the transformed solution as well as the transformed potential emerge in
explicit form. A particularly interesting application of our Darboux
transformation concerns the two-dimensional Dirac equation in
cylindrical coordinates. This equation will be derived in section 3 and
shown to admit our Darboux transformation in each of its variables
(section 4). In section 5 we state two examples of how the Darboux
transformation can be applied to the two-dimensional Dirac equation in
cylindrical coordinates. Starting from the force-free case, we obtain a
Dirac equation minimally coupled to a magnetic field, together with a
position-dependent mass.

\section{Construction of the Darboux transformation}
In the following we will introduce our generalized Dirac equation and construct a Darboux transformation by
means of the intertwining relation.
\paragraph{The Dirac equation.} The two-dimensional equation we will study in this work has the following form:
\begin{equation}\label{tdd}
\left(M~\partial_x+N~\partial_y+W_0 \right)~\psi=0,
\end{equation}
where the symbol $\partial$ denotes partial differentiation. The function $\psi=\psi(x,y)$ stands for the 4-component solution vector and $M=M(x,y),~N=N(x,y)$, and $W_0=W_0(x,y)$ denote
4 $\times$ 4 matrix functions. Furthermore, the function $W_0$ will have the usual form $W_0=E-V_0$, where $E$ is the constant
energy and $V_0=V_0(x,y)$ represents the potential. Observe that the (1+1) dimensional Dirac equation is a special case of
(\ref{tdd}), if we set
\begin{eqnarray}
E~=~0 \qquad M~=~-i~\sigma_1~\left(%
\begin{array}{cc}
  0 & I_2 \\
  I_2 & 0 \\
\end{array}%
\right) \qquad N~=~\sigma_2\left(%
\begin{array}{cc}
  0 & I_2 \\
  I_2 & 0 \\
\end{array}%
\right),
\end{eqnarray}
where $\sigma_1$, $\sigma_2$ stand for the Pauli matrices, and $I_2$ denotes the 2 $\times$ 2 identity matrix. Furthermore,
under this reduction $y$ assumes the role of the time variable.
\paragraph{The Darboux transformation.} We will now construct the Darboux transformation for our Dirac equation (\ref{tdd}). To this end, let us first multiply the equation
by $N^{-1}$ on both sides, which will simplify our subsequent calculations. After the multiplication, equation (\ref{tdd}) takes the form
\begin{eqnarray}
\left(F~\partial_x+\partial_y+U_0 \right)~\psi &=& 0, \label{tdd0}
\end{eqnarray}
where $F=N^{-1}~M$ and $U_0 = N^{-1}~W_0$. Now consider another Dirac equation of the form (\ref{tdd0}), but for a different
potential term $U_1=U_1(x,y)$:
\begin{eqnarray}
\left(F~\partial_x+\partial_y+U_1 \right)~\phi &=& 0, \label{tdd1}
\end{eqnarray}
where $\phi=\phi(x,y)$ stands for the solution. Our goal is to map solutions of (\ref{tdd0}) onto solutions of (\ref{tdd1})
by means of a linear, first-order differential operator $L$:
\begin{eqnarray}
L &=& A~\partial_x+A~B. \label{inter}
\end{eqnarray}
Here the functions $A=A(x,y)$ and $B=B(x,y)$ must be determined in such a way that the solutions $\psi$ and $\phi$ of the Dirac
equations (\ref{tdd0}) and (\ref{tdd1}), respectively, are related to each other via
\begin{eqnarray}
\phi &=& L~\psi. \nonumber
\end{eqnarray}
This is true if $L$ satisfies the following operator equation, called intertwining relation:
\begin{eqnarray}
\left(F~\partial_x+\partial_y+U_1 \right)~L &=& L~\left(F~\partial_x+\partial_y+U_0 \right). \label{interr}
\end{eqnarray}
A solution $L$ of this relation is called a Darboux operator or intertwiner for equations (\ref{tdd0}) and (\ref{tdd1}). Now, in order to
find an intertwiner of the form (\ref{inter}), we substitute its explicit form into relation (\ref{interr}) and expand both sides. Starting
with the left hand side, we obtain the following expression:
\begin{eqnarray}
\left(F~\partial_x+\partial_y+U_1 \right)~L &=& \left(F~\partial_x+\partial_y+U_1 \right)~ \left(A~\partial_x+A~B\right) \nonumber \\
&=& F~A~\partial_{xx}+A~\partial_{xy}+\left(F~A_x+F~A~B+A_y+U_1~A \right)~\partial_x+ \nonumber \\
&+& A~B~\partial_y +
F~A_x~B+F~A~B_x+A_y~B+A~B_y+U_1~A~B. \label{links}
\end{eqnarray}
In the same fashion we expand the right hand side of (\ref{interr}):
\begin{eqnarray}
L~\left(F~\partial_x+\partial_y+U_0 \right) &=& \left(A~\partial_x+A~B\right)~\left(F~\partial_x+\partial_y+U_0 \right) \nonumber \\
&=& A~F~\partial_{xx}+A~\partial_{xy}+\left(A~F_x+A~U_0+A~B~F \right)~\partial_x+ \nonumber \\
&+& A~B~\partial_y +
A~(U_0)_x+A~B~U_0. \label{rechts}
\end{eqnarray}
Our intertwining relation (\ref{interr}) is fulfilled if the expressions on its left hand side (\ref{links}) and on its right hand side
(\ref{rechts}) are the same, that is, if the coefficients of the respective derivative operators are equal on both sides. Comparison of
these coefficients gives the following conditions:
\begin{eqnarray}
F~A &=& A~F \label{con1} \\
F~A_x+F~A~B+A_y+U_1~A &=& A~F_x+A~U_0+A~B~F \label{con2} \\
F~A_x~B+F~A~B_x+A_y~B+A~B_y+U_1~A~B &=& A~(U_0)_x+A~B~U_0. \label{con3}
\end{eqnarray}
We will now solve these equations for the functions $A$, $B$ and for the transformed potential $U_1$. The first equation
(\ref{con1}) is fulfilled if $A$ and $F$ commute, a fact that we will use for the simplification of terms in subsequent calculations.
The second equation (\ref{con2}) can be solved with respect to the transformed potential $U_1$:
\begin{eqnarray}
U_1 &=& A \left(F_x+U_0+B~F-F~A^{-1}~A_x-F~B-A^{-1}A_y \right) A^{-1}. \label{pott}
\end{eqnarray}
It remains to solve the last condition (\ref{con3}). To this end, we first redefine the function $B$ that appears in the intertwiner
(\ref{inter}) as follows:
\begin{eqnarray}
B &=& -u_x~u^{-1}, \label{bu}
\end{eqnarray}
introducing a 4 $\times$ 4 matrix $u=u(x,y)$, the meaning of which will be determined below. Next, we insert the latter redefinition of $B$ and
the form (\ref{pott}) of the transformed potential into our condition (\ref{con3}):
\begin{eqnarray}
A & & \hspace{-.9cm}(U_0)_x-A~u_x~u^{-1}~U_0 = -F~A_x~u_x~u^{-1}-F~A~(u_x~u^{-1})_x-A_y~u_x~u^{-1}-A~(u_x~u^{-1})_y+ \nonumber \\
&+& \left(-A~F_x-A~U_0+A~u_x~u^{-1}~F+A~F~A^{-1}~A_x+A~F~u_x~u^{-1}+ A_y \right)~u_x~u^{-1}. \nonumber
\end{eqnarray}
This equation can be simplified by cancelling equal terms and factors on both sides, and by using (\ref{con1}). After regrouping
terms, we arrive at the following form:
\begin{eqnarray}
(U_0)_x -u_x~u^{-1}~U_0 +F_x~u_x~u^{-1}+F~u_{xx}~u^{-1}+F~u_x~(u^{-1})_x+u_{xy}~u^{-1}- \nonumber \\
- u_x~u^{-1}~F~u_x~u^{-1}+
F~u_x~u^{-1}~u_x~u^{-1} +u_x~(u^{-1})_y+U_0~u_x~u^{-1}=0. \label{inte0}
\end{eqnarray}
We will now show that this equation can be integrated with respect to the variable $x$, which yields
\begin{eqnarray}
u^{-1}~\left(F~u_x+u_y+U_0~u \right) &=& C, \label{inte}
\end{eqnarray}
where $C=C(y)$ is an arbitrary constant of integration. Let us now differentiate (\ref{inte}) with respect to $x$ and obtain equation
(\ref{inte0}). Clearly, differentiation of the right hand side of (\ref{inte}) gives zero, such that we only need to consider the
right hand side:
\begin{eqnarray}
\partial_x \Big(u^{-1}~\left(F~u_x+u_y+U_0~u \right) \Big) &=& (u^{-1})_x~F~u_x+u^{-1}~F_x~u_x+u^{-1}F~u_{xx}+(u^{-1})_x~u_y
+\nonumber \\
&+& u^{-1}~u_{xy}+(u^{-1})_x~U_0~u+u^{-1}(U_0)_x~u+u^{-1}~U_0~u_x. \nonumber
\end{eqnarray}
Multiplication by $u$ from the left and by $u^{-1}$ from the right yields
\begin{eqnarray}
u~\partial_x \Big(u^{-1}~(F~u_x \hspace{-.25cm} &+& \hspace{-.25cm} u_y+U_0~u ) \Big)~u^{-1} ~~=~~ u~(u^{-1})_x~F~u_x~u^{-1}+F_x~u_x~u^{-1}+F~u_{xx}~u^{-1}+\nonumber \\
&+& u~(u^{-1})_x~u_y~
u^{-1}+u_{xy}~u^{-1}+u~(u^{-1})_x~U_0+(U_0)_x+U_0~u_x~u^{-1}. \label{inte1}
\end{eqnarray}
Next, we make use of the following identity that holds for the derivative of an inverse matrix:
\begin{eqnarray}
(u^{-1})_x &=& -u^{-1}~u_x~u^{-1}.\nonumber
\end{eqnarray}
Clearly, this is also true for the partial derivative with respect to $y$, if $x$ is replaced by $y$. Now we apply this rule to the
derivative of the inverse matrices $u^{-1}$ in (\ref{inte1}) and obtain
\begin{eqnarray}
u~\partial_x \Big(u^{-1}~(F~u_x \hspace{-.25cm} &+& \hspace{-.25cm} u_y+U_0~u ) \Big)~u^{-1} ~=~
-u_x~u^{-1}~F~u_x~u^{-1}+F_x~u_x~u^{-1}+F~u_{xx}~u^{-1}-\nonumber \\
&-& u_x~u^{-1}~u_y~u^{-1}+u_{xy}~u^{-1}-u_x~u^{-1}~U_0+(U_0)_x+U_0~u_x~u^{-1}. \label{inte2}
\end{eqnarray}
Finally, on taking into account that
\begin{eqnarray}
-u_x~u^{-1}~u_y~u^{-1} &=& u_x~(u^{-1})_y, \nonumber
\end{eqnarray}
it is immediate to see that the left hand side of (\ref{inte0}) and the right hand side of (\ref{inte2}) coincide. This shows that our
last condition (\ref{con3}) can be integrated and turns into (\ref{inte}), which can be written in the form
\begin{eqnarray}
F~u_x+u_y+U_0~u -u~C &=& 0. \label{aux}
\end{eqnarray}
Note that this equation turns into a matrix version of our Dirac equation (\ref{tdd0}) if we choose $C=0$. At this point the meaning of
the matrix $u$ becomes clear: each of its four columns contains a solution of the auxiliary equation (\ref{aux}). Thus, in order
to perform our Darboux transformation (\ref{darboux}), we need a solution of the Dirac equation (\ref{tdd0}) and four linearly independent
solutions of the auxiliary equation (\ref{aux}). Now, we have completely resolved our
conditions (\ref{con1})-(\ref{con3}) and we have found the intertwiner (\ref{inter}) that solves our intertwining relation
(\ref{interr}):
\begin{eqnarray}
L &=& A~\left(\partial_x-u_x~u^{-1}\right). \label{l}
\end{eqnarray}
Consequently, for solutions $\psi$ and $\phi$ of our Dirac equations (\ref{tdd0}) and (\ref{tdd1}), respectively, we
have the desired mapping property $L \psi = \phi$. Next, we determine the transformed potential $U_1$ from its form
(\ref{pott}). On inserting the definition (\ref{bu}), we obtain after simplification the following expression:
\begin{eqnarray}
U_1 &=&  A \left(F_x+U_0-u_x~u^{-1}~F-F~A^{-1}~A_x+F~u_x~u^{-1}~-A^{-1}A_y \right) A^{-1}. \label{pottx}
\end{eqnarray}
Now observe that the function $A$ does not fulfill any constraint except that it must commute with $F$.
We can therefore assume $A$ to be a multiple of the 4 $\times$ 4 identity matrix $I_4$, that is, $A = a~I_4$, where $a$ is a constant that can
serve e.g. for normalization purposes. In this case, the derivatives of $A$ vanish and (\ref{pott}) simplifies to
\begin{eqnarray}
U_1 &=& A \left(F_x+U_0-u_x~u^{-1}~F+F~u_x~u^{-1} \right) A^{-1} \nonumber \\[1ex]
&=& U_0 +F_x + \Big[F,~u_x~u^{-1} \Big], \label{tpot}
\end{eqnarray}
where $[\cdot,\cdot]$ denotes the commutator. Hence, the solutions $\psi$ and $\phi$ of our Dirac equations (\ref{tdd0}) and (\ref{tdd1}), respectively, are interrelated by means of
\begin{eqnarray}
\phi~~=~~L \psi ~~=~~A~\left(\psi_x-u_x~u^{-1}~\psi \right), \label{darboux}
\end{eqnarray}
where the transformed potential $U_1$ is displayed by (\ref{tpot}).


\section{The Dirac equation in cylindrical coordinates}
We will now review how the stationary Dirac equation in three dimensions can be written in cylindrical coordinates and
how it can be reduced to a two-dimensional equation \cite{strange}. Afterwards, we show that our Darboux transformation
is applicable.
\subsection{Introduction of cylindrical coordinates} Let us start at the stationary Dirac equation in three dimensions:
\begin{eqnarray}
\Big(i~(\alpha_1~\partial_x+\alpha_2~\partial_y+\alpha_3~\partial_z)-\beta~m+E-V_0) \Big)~\Psi &=& 0,\label{D3D}
\end{eqnarray}
where the energy $E$ is a real constant, $V_0=V_0(x,y,z)$ stands for the potential, the positive constant $m$ denotes the mass, and
$\Psi=\Psi(x,y,z)$ represents the 4-component solution. Furthermore, recall that the matrices $\alpha_j$, $j=1,2,3$, and
$\beta$ are given by
\begin{eqnarray}
\alpha_j ~~=~~ \left(\begin{array}{lll}
0 & \sigma_j \\
\sigma_j & 0
\end{array}
\right) \qquad \beta ~~=~~ \left(\begin{array}{lll}
I_2 & 0 \\
0 & -I_2
\end{array}
\right), \label{alphabeta}
\end{eqnarray}
where $\sigma_j$, $j=1,2,3$, denote the usual Pauli matrices. Now we introduce the change from cartesian to
cylindrical coordinates $(x,y,z) \mapsto (\rho,\varphi,z)$ by means of the usual relations
\begin{eqnarray}
x ~=~\rho~\cos(\varphi) \qquad y ~=~\rho~\sin(\varphi) \qquad z~=~z. \label{coordsystem}
\end{eqnarray}
Furthermore we suppose that the solution $\Psi$ of our Dirac equation (\ref{D3D}) depends on $\rho$ and $\varphi$, while its
dependence on $z$ can be separated off via
\begin{eqnarray}
\label{sol18} \Psi&=&\exp(i~p_z~z)~\psi,
\end{eqnarray}
where $\psi=\psi(\rho,\varphi)$ and $p_z$ stands for the eigenvalue of the momentum operator $p_z=-i~\partial_z$.
Next, we substitute (\ref{sol18}) into our Dirac equation (\ref{D3D}) and rewrite the partial derivatives in cylindrical coordinates,
which renders the Dirac equation in the following form \cite{strange}:
\begin{eqnarray}
\label{canal}
\Bigg(\beta_1~\partial_\rho+\frac{\beta_2}{\rho}~\partial_\varphi-p_z~\beta_3-\beta~m+E-V_0\Bigg)~\psi=0,
\end{eqnarray}
where the matricial coefficients are given as follows:
\begin{eqnarray}
\beta_1 &=&
\left( \begin{array}{ccccc}
0 & 0 & 0 & \exp(-i~\varphi) \\
0 & 0 & \exp(i~\varphi) & 0 \\
0 & \exp(-i~\varphi) & 0 & 0 \\
\exp(i~\varphi) & 0 & 0 & 0
\end{array}
\right) \label{beta1} \\[2ex]
\beta_2 &=&
\left( \begin{array}{ccccc}
0 & 0 & 0 & -i~\exp(-i~\varphi) \\
0 & 0 & i~\exp(i~\varphi) & 0 \\
0 & -i~\exp(-i~\varphi) & 0 & 0 \\
i~\exp(i~\varphi) & 0 & 0 & 0
\end{array}
\right) \label{beta2} \\[2ex]
\beta_3 &=&
\left( \begin{array}{ccccc}
0 & 0 & 1 & 0 \\
0 & 0 & 0 & -1 \\
1 & 0 & 0 & 0 \\
0 & -1 & 0 & 0
\end{array}
\right). \label{beta3}
\end{eqnarray}
It is immediate to see that our Dirac equation in the form (\ref{canal}) involves only the two variables $\rho$ and $\varphi$, and that
it is now a special case of our two-dimensional equation (\ref{tdd}).


\subsection{The Darboux transformation in cylindrical coordinates}
Our goal is to make the Darboux transformation (\ref{darboux}) applicable to the Dirac equation in cylindrical coordinates
(\ref{canal}), such that we obtain solutions $\phi$ and associated potentials $V_1$ for the equation
\begin{eqnarray}
\Bigg(\beta_1~\partial_\rho+\frac{\beta_2}{\rho}~\partial_\varphi-p_z~\beta_3-\beta~m+E-V_1\Bigg)~\phi=0. \label{canalx}
\end{eqnarray}
The first step in applying our Darboux transformation is to convert the Dirac equation in cylindrical coordinates into the
form (\ref{tdd0}). Since there are two terms containing partial derivatives on the left hand side of (\ref{canal}), we have
two possibilities of arriving at the desired form of our Dirac equation. Consequently, we end up with two different
Darboux transformations.
\paragraph{First Darboux transformation.} We multiply our equation (\ref{canal}) by $\beta_1^{-1}$ from the left:
\begin{eqnarray}
\Bigg(\partial_\rho+\frac{\beta_1^{-1}~\beta_2}{\rho}~\partial_\varphi-p_z~\beta_1^{-1}~\beta_3-\beta_1^{-1}~\beta~m
+\beta_1^{-1}~(E-V_0)\Bigg)~\psi=0, \label{canal1}
\end{eqnarray}
Now we introduce the abbreviations
\begin{eqnarray}
\gamma ~=~ \frac{\beta_1^{-1}~\beta_2}{\rho} \qquad U_0 ~=~-p_z~\beta_1^{-1}~\beta_3-\beta_1^{-1}~\beta~m
+\beta_1^{-1}~(E-V_0), \label{gammau00}
\end{eqnarray}
which render equation (\ref{canal1}) in the form
\begin{eqnarray}
\Bigg(\partial_\rho+\gamma~\partial_\varphi+U_0\Bigg)~\psi=0. \label{canal2}
\end{eqnarray}
On comparing this equation with our Dirac equation (\ref{tdd0}), we see that both coincide if in the latter equation we choose
$F=\gamma$ and if we identify the variable $x$ with $\varphi$ and $y$ with $\rho$, respectively. Thus, our Darboux transformation
(\ref{darboux}) becomes applicable. Let $u=u(\rho,\varphi)$ be a solution of the following auxiliary equation
\begin{eqnarray}
u_\rho+\gamma~u_\varphi+U_0~u -u~C &=& 0, \label{aux1}
\end{eqnarray}
where $C=C(\rho)$ is an arbitrary function. Clearly, this auxiliary equation is just (\ref{aux}) with the same identifications of
parameter and variables that has been done for (\ref{canal2}). Now, the function $\phi=\phi(\rho,\varphi)$ defined as
\begin{eqnarray}
\phi ~~=~~ L~\psi ~~=~~A~\left(\psi_\varphi-u_\varphi~u^{-1}~\psi \right), \label{darbouxc}
\end{eqnarray}
is a solution of the Dirac equation
\begin{eqnarray}
\Bigg(\partial_\rho+\gamma~\partial_\varphi+U_1\Bigg)~\phi=0, \label{canal3}
\end{eqnarray}
where the transformed potential $U_1$ is given by (\ref{pottx}), which reads in the present case
\begin{eqnarray}
U_1 &=&  A \left(\gamma_\varphi+U_0-u_\varphi~u^{-1}~\gamma-\gamma~A^{-1}~A_\varphi+\gamma~u_\varphi~u^{-1}~-A^{-1}A_\rho \right)
A^{-1} \nonumber \\
&=& A \left(U_0-\frac{u_\varphi~u^{-1}~\beta_1^{-1}~\beta_2}{\rho}-\frac{\beta_1^{-1}~\beta_2~A^{-1}~A_\varphi}{\rho}
+\frac{\beta_1^{-1}~\beta_2~u_\varphi~u^{-1}}{\rho}
-A^{-1}A_\rho \right) A^{-1},\nonumber \\  \label{pottx1}
\end{eqnarray}
where in the last step we made use of the fact that $\gamma_\varphi=0$. This expression can be simplified further, if we
assume $A$ to be a multiple of a constant $a$, that is, $A=a I_4$, where $I_4$ stands for the 4 $\times$ 4 identity matrix. Recall that
this assumption is valid, as the only constraint $A$ must fulfill is to commute with $\gamma$. The transformed potential (\ref{pottx1})
then takes the following form (recall $\gamma_\varphi=0$)
\begin{eqnarray}
U_1 &=& U_0 +\gamma_\varphi + \Big[\gamma,~u_\varphi~u^{-1} \Big] \nonumber \\
&=& U_0 +\frac{1}{\rho}~\Big[\beta_1^{-1}~\beta_2,~u_\varphi~u^{-1} \Big]. \label{potdf1}
\end{eqnarray}
Thus, the solutions $\psi$ and $\phi$ of our Dirac equations (\ref{canal2}) and (\ref{canal3}) in cylindrical coordinates, respectively,
are interrelated via our Darboux transformation (\ref{darbouxc}), where the auxiliary solution $u$ and the transformed potential $U_1$
fulfill (\ref{aux1}) and (\ref{pottx1}), respectively. Note that the potential difference (\ref{potdf1}) does not involve the actual potentials $V_0$ and $V_1$, as they occur in the
Dirac equation (\ref{canal}) and in its Darboux-transformed counterpart (\ref{canalx}). The Dirac equation in its
initial form (\ref{canal}) obtained the form (\ref{canal2}) through a multiplication by $\beta_1^{-1}$. The inverse of this we now
multiply by (\ref{potdf1}) and make use of the relation (\ref{gammau00}):
\begin{eqnarray}
V_1 &=& V_0
+\frac{\beta_1}{\rho}~\Big[\beta_1^{-1}~\beta_2,u_\varphi~u^{-1}\Big].
\label{pottx5}
\end{eqnarray}
This is the final expression for the potential difference, where the potentials $V_1$ and $V_0$ are associated with the initial and the
final Dirac equation (\ref{canal}) and (\ref{canalx}), respectively.
\paragraph{Second Darboux transformation.} Let us now return to our Dirac equation it its form (\ref{canal}), and let us
multiply this equation from the left by
$\rho~\beta_2^{-1} = \rho~\beta_2$:
\begin{eqnarray}
\Bigg(\rho~\beta_2~\beta_1~\partial_\rho+\partial_\varphi-p_z~\rho~\beta_2~\beta_3-\rho~\beta_2~\beta~m+\rho~\beta_2~(E-V_0)\Bigg)~\psi=0,
\label{canal4}
\end{eqnarray}
Similar to the previous case, we introduce the following abbreviations:
\begin{eqnarray}
\gamma ~=~\rho~\beta_2~\beta_1 \qquad U_0 ~=~ -p_z~\rho~\beta_2~\beta_3-\rho~\beta_2~\beta~m+\rho~\beta_2~(E-V_0), \label{gammau0}
\end{eqnarray}
which turn (\ref{canal4}) into the equation
\begin{eqnarray}
\Bigg(\gamma~\partial_\rho+\partial_\varphi+U_0\Bigg)~\psi=0.
\label{canal5}
\end{eqnarray}
Similar to our previous case (\ref{canal2}), we now compare our general Dirac equation (\ref{tdd0}) with (\ref{canal5}). These equations
coincide, if we have $F=\gamma$ and if we identify the variables $\rho$ and $\varphi$ with $x$ and $y$, respectively. In this case we
can apply our Darboux transformation (\ref{darboux}), the auxiliary function $u=u(\rho,\varphi)$ for which must be a solution
of the equation
\begin{eqnarray}
\gamma~u_\rho+u_\varphi+U_0~u -u~C &=& 0, \label{aux2}
\end{eqnarray}
where $C=C(\varphi)$ is an arbitrary function. Then, as in the previous case, the function $\phi=\phi(\rho,\varphi)$ defined by
\begin{eqnarray}
\phi ~~=~~ L~\psi ~~=~~A~\left(\psi_\rho-u_\rho~u^{-1}~\psi \right), \label{darbouxc1}
\end{eqnarray}
is a solution of the Dirac equation
\begin{eqnarray}
\Bigg(\gamma~\partial_\rho+\partial_\varphi+U_1\Bigg)~\phi=0. \label{canal6}
\end{eqnarray}
The transformed potential $U_1$ is given as the following particular case of (\ref{pottx}):
\begin{eqnarray}
U_1 &=&  A \left(\gamma_\rho+U_0-u_\rho~u^{-1}~\gamma-\gamma~A^{-1}~A_\rho+\gamma~u_\rho~u^{-1}~-A^{-1}A_\rho \right)
A^{-1} \nonumber \\
&=& A \left(\beta_2~\beta_1+U_0-\rho^2~u_\rho~u^{-1}~\beta_2~\beta_1-\rho~\beta_2~\beta_1~A^{-1}~A_\rho+\rho~\beta_2~\beta_1
~u_\rho~u^{-1}\hspace{-.1cm}-A^{-1}A_\rho \right), \nonumber \\ \label{pottx2}
\end{eqnarray}
In order to simplify this expression, let us suppose $A=a I_4$ for a constant $a$. The transformed potential (\ref{pottx2})
then takes the following form
\begin{eqnarray}
U_1 &=& U_0 +\gamma_\rho + \Big[\gamma,~u_\rho~u^{-1} \Big] \nonumber \\
&=& U_0 +\beta_2~\beta_1+\rho~\Big[\beta_2~\beta_1,~u_\rho~u^{-1} \Big]. \label{pottxxx}
\end{eqnarray}
Hence, in the present case the solutions $\psi$ and $\phi$ of our Dirac equations (\ref{canal5}) and (\ref{canal6})
in cylindrical coordinates, respectively, are interrelated via our Darboux transformation (\ref{darbouxc1}),
where the auxiliary function $u$ and the transformed potential satisfy (\ref{aux2}) and (\ref{pottx2}), respectively.
Let us finally remark that again relation (\ref{pottxxx}) can be converted  into a relation between the actual potentials $V_0$ and $V_1$ in the Dirac equation
(\ref{canal}) and its Darboux-transformed counterpart, respectively. To this end, note that the Dirac equation in its
initial form (\ref{canal}) obtained the form (\ref{canal5}) through a multiplication by $\rho~\beta_2$. The inverse of this we now
multiply by (\ref{pottxxx}) and make use of the relation (\ref{gammau0}):
\begin{eqnarray}
V_1
&=&V_0+\frac{\beta_1}{\rho}+\beta_2~\Big[\beta_2~\beta_1,u_\rho~u^{-1}\Big].
\label{pot2end}
\end{eqnarray}
This is the final expression for the transformed potential $V_1$ associated with the Dirac equation in its form (\ref{canalx}).

\section{Examples}
 We will now apply our Darboux transformation to the Dirac
equation in cylindrical coordinates (\ref{canal}), for the sake of
simplicity we consider the force-free case, that is, $V_0=0$. To
this end, let us first break up our Dirac equation into two
components. We write its solution in the form $\psi =
(\psi_1,\psi_2)^T$, where each of the functions $\psi_1$ and
$\psi_2$ stands for a two-component vector. Next, we need the
following submatrices of the $\beta_j$, $j=1,2,3$, as given in
(\ref{beta1})-(\ref{beta3}):
\begin{eqnarray}
\tau_1 = \left(
\begin{array}{cccc}
0 & \exp(-i \varphi) \\
\exp(i \varphi) & 0
\end{array}
\right) \quad \hspace{-.15cm} \tau_2 = \left(
\begin{array}{cccc}
0 & -i~\exp(-i \varphi) \\
i~\exp(i \varphi) & 0
\end{array}
\right) \quad \hspace{-.15cm}
\tau_3 = \left(
\begin{array}{cccc}
1 & 0 \\
0 & -1
\end{array}
\right) \hspace{-.15cm}. \nonumber \\
\label{taus}
\end{eqnarray}
Now we are ready to decompose the Dirac equation (\ref{canal})
into two equations, each of which has two components.
\begin{eqnarray}
\tau_1~(\psi_1)_\rho+\frac{\tau_2}{\rho}~(\psi_1)_\varphi-p_z~\tau_3~\psi_1+(E+m)~\psi_2 &=& 0\nonumber \\[1ex]
\tau_1~(\psi_2)_\rho+\frac{\tau_2}{\rho}~(\psi_2)_\varphi-p_z~\tau_3~\psi_2+(E-m)~\psi_1 &=& 0. \nonumber
\end{eqnarray}
Next, in order to apply our Darboux transformation
(\ref{darboux}), we need four linearly independent solutions of
our auxiliary equation (\ref{aux}) for $V_0=0$, which will form
the columns of our auxiliary matrix $u$. From now on we write this
matrix $u$ in 2 $\times$ 2 components as
\begin{eqnarray}
u &=& \left(
\begin{array}{ll}
u_1 & u_2 \\ u_3 & u_4
\end{array}
\right). \nonumber
\end{eqnarray}
For the sake of simplicity let us assume that
the arbitrary matrix $C=C(\varphi)$ that appears on the left hand side of our auxiliary equation (\ref{aux}), has the following form:
\begin{eqnarray}
 C&=& \left(%
\begin{array}{cc}
  C_1~I_2 & 0 \\
  0 & C_4~I_2 \\
\end{array}%
\right), \label{auxc}
\end{eqnarray}
where each entry contains a 2 $\times$ 2 matrix, and both matrices $C_1$ and $C_4$ can depend on $\varphi$. Let us now substitute (\ref{auxc}) and the matrix $u$ into the auxiliary equation
(\ref{aux}), which in cylindrical coordinates then splits up into the following four 2 $\times$ 2 matrix components:
\begin{eqnarray}
\label{S1}\tau_1 ~(u_3)_\rho+\frac{\tau_2}{\rho}~(u_3)_\varphi-p_z~\tau_3~u_3+(E+m)~u_1&=&\frac{\tau_2~C_1}{\rho}~u_3\\
\label{S2}\tau_1~(u_1)_\rho+\frac{\tau_2}{\rho}~(u_1)_\varphi-p_z~\tau_3~u_1+(E-m)~u_3&=&\frac{\tau_2~C_1}{\rho}~u_1\\
\label{S3}\tau_1~(u_4)_\rho+\frac{\tau_2}{\rho}~(u_4)_\varphi-p_z~\tau_3~u_4+(E+m)~u_2&=&\frac{\tau_2~C_4}{\rho}~u_4 \\
\label{S4}\tau_1~(u_2)_\rho+\frac{\tau_2}{\rho}~(u_2)_\varphi-p_z~\tau_3~u_2+(E-m)~u_4&=&\frac{\tau_2~C_4}{\rho}~u_2.
\end{eqnarray}
We will now find two solutions of this system of equations, each of which leads to the generation of a new solvable
Dirac potential in cylindrical coordinates.

\subsection{First example}
In our first example we take equations \eqref{S1} and \eqref{S2} in the case $E=m$:
\begin{eqnarray}
 \label{S1E=m}\tau_1~(u_3)_\rho-p_z~\tau_3~u_3+2~m~u_1&=& \frac{\tau_2}{\rho}~\bigg(C_1~u_3-(u_3)_\varphi \bigg)\\
\label{S2E=m} \tau_1~(u_1)_\rho-p_z~\tau_3~u_1&=& \frac{\tau_2}{\rho}~\bigg(C_1~u_1-(u_1)_\varphi \bigg).
\end{eqnarray}
In order to simplify these two equation further, let us now assume that $C_1= (u_3)_\varphi=0$. We obtain the following
equations:
\begin{eqnarray}
\label{S1E=mG1=0}\tau_1~(u_3)_\rho-p_z~\tau_3~u_3+2~m~u_1&=&0,\\
\label{S2E=mG1=0} \tau_1~(u_1)_\rho-p_z~\tau_3~u_1&=&-\frac{\tau_2}{\rho}~(u_1)_\varphi.
\end{eqnarray}
Now we solve the first of these equations with respect to $u_1$ and substitute the result into the second equation. The
first equation yields
\begin{equation}\label{u1}
    u_1=\frac{p_z~\tau_3~u_3-\tau_1~ (u_3)_\rho}{2~m},
\end{equation}
the partial derivatives of which read
\begin{eqnarray}
(u_1)_\rho&=&\frac{p_z~\tau_3~(u_3)_\rho-\tau_1~ (u_3)_{\rho\rho}}{2~m} \label{u1rho}  \\
(u_1)_\varphi&=&-\frac{\tau_2~ (u_3)_\rho}{2~m} \label{u1phi}.
\end{eqnarray}
If we now replace $u_1$ in \eqref{S2E=mG1=0} by inserting our findings (\ref{u1}), (\ref{u1rho}) and  (\ref{u1phi}), we get
\begin{eqnarray}\label{equationforu3}
-(u_3)_{\rho\rho}-p_z^2~u_3&=&\frac{1}{\rho}~(u_3)_\rho
\end{eqnarray}
Let us suppose $p_z=0,$ then this becomes
\begin{eqnarray}\label{equationforu3forpz=0}
-(u_3)_{\rho\rho}&=&\frac{1}{\rho}~(u_3)_\rho. \nonumber
\end{eqnarray}
A particular solution $u_3$ of this equation is given by
\begin{eqnarray}
u_3&=&\log(\rho)~I_2. \nonumber
\end{eqnarray}
Now that we have found $u_3$, we can construct $u_1$ from relation (\ref{u1}):
\begin{eqnarray}
u_1 &=& -\frac{\tau_1}{2~m~\rho}. \nonumber
\end{eqnarray}
It remains to find the matrix functions $u_2$ and $u_4$. To this end, we proceed in a similar way as done in the previous calculation.
First we rewrite equations \eqref{S3} and \eqref{S4} for the case $E=-m$, $C_4=0$, $(u_2)_\varphi=0$, and $p_z=0$:
\begin{eqnarray}
\tau_1~(u_4)_\rho&=&-\frac{\tau_2}{\rho}~ (u_4)_\varphi,\label{S3aftersuppos} \\
\tau_1~(u_2)_\rho&=&2~m~u_{4}.\label{S4aftersuppos}
\end{eqnarray}
The second equation can now be solved for $u_4$:
\begin{eqnarray}
u_4&=&\frac{\tau_1}{2~m}~(u_2)_\rho. \nonumber
\end{eqnarray}
The partial derivatives of this function read as follows:
\begin{eqnarray}
(u_4)_\rho&=&\frac{\tau_1}{2~m}~(u_2)_{\rho\rho} \nonumber \\
(u_4)_\varphi&=&\frac{\tau_2}{2~m}~(u_2)_\rho. \nonumber
\end{eqnarray}
Now we insert the function $u_4$ and its partial derivatives into the remaining equation (\ref{S3aftersuppos}):
\begin{eqnarray}
u_2&=&\log(\rho)~I_2. \nonumber
\end{eqnarray}
As in the previous case, this is a particular solution. Now that we have found the four components of our matrix $u$, we can state
this matrix in its explicit form:
\begin{equation}\label{matrixu}
u=\left(%
\begin{array}{ll}
 {\displaystyle{-\frac{\tau_1}{2~m~\rho}}} & \log(\rho)~I_2 \\[2ex]
  \log(\rho)~I_2 &{\displaystyle{\frac{\tau_1}{2~m~\rho}}} \\
\end{array}
\right).
\end{equation}
In order to find the transformed potential, the general form of which is displayed in (\ref{pottx2}), we need the
partial derivative $u_\rho$ and the inverse $u^{-1}$ of $u$:
\begin{eqnarray}
u_{\rho}&=&\left(%
\begin{array}{cc}
{\displaystyle{\frac{\tau_1}{2~m~\rho^2}}} & {\displaystyle{\frac{1}{\rho}~I_2}} \\[3ex]
{\displaystyle{\frac{1}{\rho}~I_2}} &{\displaystyle{-\frac{\tau_1}{2~m~\rho^2}}}
\end{array} \label{urho}
\right) \nonumber \\[2ex]
u^{-1}&=&\frac{1}{\log^2(\rho)+{\displaystyle{\frac{1}{4m^2\rho^2}}}}\left(%
\begin{array}{cc}
{\displaystyle{-\frac{\tau_1}{2~m~\rho}}} & \log(\rho)~I_2 \\
  \log(\rho)~I_2 &{\displaystyle{\frac{\tau_1}{2~m~\rho}}} \\
\end{array}%
\right).
\end{eqnarray}
Multiplication of the latter two matrices gives
\begin{eqnarray}
u_{\rho}~u^{-1}=\frac{1}{\log^2(\rho)+{\displaystyle{\frac{1}{4~m^2~\rho^2}}}}\left(%
\begin{array}{ll}
 {\displaystyle{\left(\frac{\log(\rho)}{\rho}-\frac{1}{4~m^2~\rho^3}\right)~I_2}} & {\displaystyle{-\frac{\tau_1}{2~m~\rho^2}~\bigg(1+\log(\rho)\bigg)}} \\[2ex]
 {\displaystyle{ \frac{\tau_1}{2~m~\rho^2}~\bigg(1+\log(\rho)\bigg)}} &{\displaystyle{\left(\frac{\log(\rho)}{\rho}-\frac{1}{4~m^2~\rho^3}\right)~I_2}}
\end{array}%
\right) \nonumber \\ \label{urhou-1}
\end{eqnarray}
We are now in position to construct the new solvable Dirac potential $V_1$. To this end, we must evaluate relation
(\ref{pot2end}) for the present case, which for the sake of convenience we repeat here, note that we have $V_1=0$:
\begin{eqnarray}
V_1 &=&
\frac{\beta_1}{\rho}+\beta_2~\Big[\beta_2~\beta_1,u_\rho~u^{-1}\Big].
\label{pot3end}
\end{eqnarray}
We start by computing the first argument of the commutator in (\ref{pot3end}), using the definitions (\ref{beta1}), (\ref{beta2})
and (\ref{taus}):
\begin{eqnarray}
\beta_2~\beta_1=\left(%
\begin{array}{cc}
  0 & \tau_2 \\
  \tau_2 & 0 \\
\end{array}%
\right)\left(%
\begin{array}{cc}
  0 & \tau_1 \\
  \tau_1 & 0 \\
\end{array}%
\right)=\left(%
\begin{array}{cc}
  \tau_2~\tau_1 & 0 \\
  0 & \tau_2~\tau_1 \\
\end{array}%
\right)=-i\left(%
\begin{array}{cc}
  \tau_3 & 0 \\
  0 & \tau_3 \\
\end{array}%
\right).
\end{eqnarray}
We will now compute the full commutator in (\ref{pot3end}). In order to avoid large expressions, we introduce the following
abbreviations:
\begin{eqnarray}
A &=& \frac{1}{\log^2(\rho)+{\displaystyle{\frac{1}{4~m^2~\rho^2}}}}~\left(\frac{\log(\rho)}{\rho}-\frac{1}{4~m^2~\rho^3} \right)~I_2 \nonumber \\
B &=& \frac{1}{\log^2(\rho)+{\displaystyle{\frac{1}{4~m^2~\rho^2}}}}~\left( \frac{1}{2~m~\rho^2}~\bigg(1+\log(\rho)\bigg) \right). \label{babb}
\end{eqnarray}
These settings render the product $u_{\rho}~u^{-1}$ in the compact form
\begin{equation}\label{D}u_{\rho}~u^{-1}=\left(%
\begin{array}{cc}
  A & -B~\tau_1 \\
  B~\tau_1 & A \\
\end{array}%
\right).
\end{equation}
Using this form of $u_{\rho}~u^{-1}$ we can finally evaluate the commutator, which reads
\begin{eqnarray}\label{calcul}
   [\beta_2~\beta_1,u_{\rho}~u^{-1}]&=&-i~\left[\left(%
\begin{array}{cc}
  \tau_3& 0 \\
  0 & \tau_3 \\
\end{array}%
\right),\left(%
\begin{array}{cc}
  A & -B~\tau_1 \\
  B~\tau_1 & A \\
\end{array}%
\right)\right] \nonumber \\
&=&-2~i~\left(%
\begin{array}{cc}
  0 & -B~\tau_3~\tau_1 \\
  B~\tau_3~\tau_1 & 0 \\
\end{array}%
\right).
\end{eqnarray}
Next, we have to multiply this by $\beta_2$ from the left:
\begin{eqnarray}
 \beta_2~[\beta_2~\beta_1,u_{\rho}~u^{-1}]&=&-2~i~\left(%
\begin{array}{cc}
  0 & \tau_2 \\
  \tau_2 & 0 \\
\end{array}%
\right)\left(%
\begin{array}{cc}
  0 & -B~\tau_3~\tau_1 \\
  B~\tau_3~\tau_1 & 0 \\
\end{array}%
\right) \\&=&-2~i~B~\tau_2~\tau_3~\tau_1\left(%
\begin{array}{cc}
I_2&0 \\
0&-I_2\\
\end{array}%
\right). \label{commutator}
\end{eqnarray}
The product $\tau_2~\tau_3~\tau_1$ turns out to be
\begin{eqnarray}
\tau_2~\tau_3~\tau_1&=& i~\left(%
\begin{array}{cc}
  0 & -\exp\left(-i~\varphi\right) \\
  \exp\left(i~\varphi \right) & 0
\end{array}%
\right)
\left(%
\begin{array}{cc}
  1 & 0 \\
  0 & -1
\end{array}%
\right)
\left(%
\begin{array}{cc}
  0 & \exp\left(-i~\varphi\right) \\
  \exp\left(i~\varphi\right) & 0
\end{array}%
\right) \nonumber \\
&=& i~I_2. \label{beta}
\end{eqnarray}
Taking this into account, the commutator (\ref{commutator}) assumes the following final form, recall the definition of $B$ as
given in (\ref{babb}):
\begin{eqnarray}
\beta_2~[\beta_2~\beta_1,u_{\rho}~u^{-1}]&=& 2~B~\beta \label{2bbeta} \\
&=&\frac{2~(1+\log(\rho))~\beta}{2~m~\rho^2~(\log^2(\rho)+\frac{1}{4~m^2~\rho^2})} \nonumber \\
&=&\frac{4~m~(1+\log(\rho))~\beta}{4~m^2~\rho^2~\log^2(\rho)+1}. \nonumber
\end{eqnarray}
Now we take the previous result, combine it with (\ref{beta1}), and substitute into (\ref{pot3end}):
\begin{eqnarray}
V_1 &=& \left( \begin{array}{ccccc}
0 & 0 & 0 & {\displaystyle{\frac{1}{\rho}}}~\exp(-i~\varphi) \\
0 & 0 & {\displaystyle{\frac{1}{\rho}}}~\exp(i~\varphi) & 0 \\
0 & {\displaystyle{\frac{1}{\rho}}}~\exp(-i~\varphi) & 0 & 0 \\
{\displaystyle{\frac{1}{\rho}}}~\exp(i~\varphi) & 0 & 0 & 0
\end{array}
\right)+ \nonumber \\[2ex]
&+& \left( \begin{array}{ccccc}
\frac{4~m~(1+\log(\rho))}{4~m^2~\rho^2~\log^2(\rho)+1} & 0 & 0 & 0 \\
0 & \frac{4~m~(1+\log(\rho))}{4~m^2~\rho^2~\log^2(\rho)+1} & 0 & 0 \\
0 & 0 & -\frac{4~m~(1+\log(\rho))}{4~m^2~\rho^2~\log^2(\rho)+1} & 0 \\
0 & 0 & 0 & -\frac{4~m~(1+\log(\rho))}{4~m^2~\rho^2~\log^2(\rho)+1}
\end{array}
\right) \nonumber \\[2ex]
&=& \left( \begin{array}{ccccc}
\frac{4~m~(1+\log(\rho))}{4~m^2~\rho^2~\log^2(\rho)+1} & 0 & 0 & {\displaystyle{\frac{1}{\rho}}}~\exp(-i~\varphi) \\
0 & \frac{4~m~(1+\log(\rho))}{4~m^2~\rho^2~\log^2(\rho)+1} & {\displaystyle{\frac{1}{\rho}}}~\exp(i~\varphi) & 0 \\
0 & {\displaystyle{\frac{1}{\rho}}}~\exp(-i~\varphi) & -\frac{4~m~(1+\log(\rho))}{4~m^2~\rho^2~\log^2(\rho)+1} & 0 \\
{\displaystyle{\frac{1}{\rho}}}~\exp(i~\varphi) & 0 & 0 &
-\frac{4~m~(1+\log(\rho))}{4~m^2~\rho^2~\log^2(\rho)+1}
\end{array}
\right). \nonumber \\ \label{endv1}
\end{eqnarray}
This is the final form of our Darboux-transformed potential $V_1$. The corresponding solution $\phi$ of the Dirac equation
(\ref{canalx}) can be constructed from the Darboux transformation (\ref{darbouxc1}).

\subsection{Second example}
We start again from the system of equations (\ref{S1})-(\ref{S4}), making the following assumptions:
\begin{eqnarray}
C_4 ~=~0 \qquad (u_3)_\varphi~=~(u_1)_\varphi~=~0 \qquad p_z~=~0. \label{simpli}
\end{eqnarray}
Furthermore, we redefine $E=\varepsilon$, together with the restriction $0<\varepsilon<m$. These settings render
equations \eqref{S3} and \eqref{S4} in the following form:
\begin{eqnarray}
\label{S1E=varepsilon}\tau_1~(u_4)_\rho+(\varepsilon+m)~u_2&=&0\\
\label{S2E=varepsilon} \tau_1~(u_2)_\rho+(\varepsilon-m)~u_4&=&0.
\end{eqnarray}
We can solve the first of these equations with respect to $u_2$:
\begin{eqnarray}\label{u111}
u_2=-\frac{\tau_1}{\varepsilon+m}~ (u_4)_\rho.
\end{eqnarray}
On inserting this into equation (\ref{S2E=varepsilon}) and introducing $k^2=\varepsilon^2-m^2$, we get
\begin{eqnarray}
(u_4)_{\rho\rho}&=&k^2~u_{4}. \label{u33}
\end{eqnarray}
It is immediate to see that a particular pair of solutions to equations (\ref{u111}) and (\ref{u33})
\begin{eqnarray}\label{solute}
u_2~=~-\frac{\tau_1~k~\cosh(k~\rho)}{\varepsilon+m} \qquad u_4~=~\sinh(k~\rho)~I_2.
\end{eqnarray}
It remains to find a pair of solutions for the equations $\eqref{S1}$ and $\eqref{S2}$. To this end, observe that under the settings
(\ref{simpli}) these equations become equivalent to $\eqref{S3}$ and $\eqref{S4}$, respectively, provided we set $C_1$=0. Thus,
we will proceed as in the previous case, but introducing the setting $E=-\varepsilon$. Note that this is allowed, since each pair of
the functions $(u_1,u_3)$ and $(u_2,u_4)$ must form a solution of our auxiliary equation (\ref{aux}), but not necessarily at the
same energy $E$. Now, on employing (\ref{simpli}), $C_1=0$ and $E=-\varepsilon$, we solve equation \eqref{S1} for
$u_3$ and substitute the result in \eqref{S2}:\begin{eqnarray}\label{u11}
u_3~=~\frac{\tau_1}{\varepsilon+m}~(u_1)_\rho \qquad (u_1)_{\rho\rho}~=~k^2~u_{1}.
\end{eqnarray}
A particular solution of these two equations is given by\begin{eqnarray}\nonumber
u_1~=~\sinh(k~\rho)~I_2 \qquad u_3~=~\frac{\tau_1~k~\cosh(k~\rho)}{\varepsilon+m}.
\end{eqnarray}
Now we have determined all four components of our matrix $u$, which reads in explicit form:
\begin{equation}\nonumber
 u~=~\left(%
\begin{array}{cc}
  \sinh(k~\rho)~I_2 & {\displaystyle{-\frac{k~\tau_1~\cosh(k~\rho)}{\varepsilon+m}}}\\
 {\displaystyle{\frac{k~\tau_1~\cosh(k~\rho)}{\varepsilon+m}}} & \sinh(k~\rho)~I_2 \\
  \end{array}%
\right)
\end{equation}
We now perform the same steps as in the previous example. First we compute the inverse $u^{-1}$ and the partial derivative
$u_\rho$ of $u$, afterwards we use these results for evaluating the transformed potential.
\begin{eqnarray}
u^{-1}&=&\frac{1}{{\displaystyle{\frac{k^2~\cosh^2(k~\rho)}{(\varepsilon+m)^2}}}+\sinh^2(k~\rho)}~\left(%
\begin{array}{cc}
  \sinh(k~\rho)~I_2 & {\displaystyle{-\frac{k~\tau_1~\cosh(k~\rho)}{\varepsilon+m}}} \\
  {\displaystyle{\frac{k~\tau_1~\cosh(k~\rho)}{\varepsilon+m}}} & \sinh(k~\rho)~I_2 \\
  \end{array}%
\right) \nonumber \\[1ex]
u_{\rho}&=&k~\left(%
\begin{array}{cc}
 \cosh(k~\rho)~I_2  & {\displaystyle{-\frac{k~\tau_1~\sinh(k~\rho)}{\varepsilon+m}}} \\
  {\displaystyle{\frac{k~\tau_1~\sinh(k~\rho)}{\varepsilon+m}}} &\cosh(k~\rho)~I_2 \\
\end{array}%
\right). \nonumber
\end{eqnarray}
The product of these two matrices appears in the transformed potential $V_1$, as given in (\ref{pot2end}), and it reads
\begin{eqnarray}\nonumber
u_{\rho}~u^{-1}=\frac{1}{\frac{k}{\varepsilon+m}\cosh^2(k~\rho)+\frac{\varepsilon+m}{k}\sinh^2(k~\rho)}\left(%
\begin{array}{cc}
  m~\sinh(2~k~\rho)~I_2& -k~\tau_1~\cosh(2~k~\rho) \\
  k~\tau_1~\cosh(2~k~\rho) & m~\sinh(2~k~\rho)~I_2 \\
\end{array}%
\right).
\end{eqnarray}
Now, in order to evaluate (\ref{pot2end}), we introduce the following abbreviations
\begin{eqnarray}
A &=& \frac{m~\sinh(2~k~\rho)}{{\displaystyle{\frac{k}{\varepsilon+m}\cosh^2(k~\rho)+\frac{\varepsilon+m}{k}\sinh^2(k~\rho)}~I_2}} \nonumber \\
B &=& \frac{k~\cosh(2~k~\rho)}{{\displaystyle{\frac{k}{\varepsilon+m}\cosh^2(k~\rho)+\frac{\varepsilon+m}{k}\sinh^2(k~\rho)}}}. \nonumber
\end{eqnarray}
These settings render our matrix product $u_{\rho} u^{-1}$, as given by (\ref{urhou-1}), in the following form:
\begin{equation}\nonumber u_{\rho}~u^{-1}=\left(%
\begin{array}{cc}
  A & -B~\tau_1 \\
  B~\tau_1 & A \\
\end{array}%
\right).
\end{equation}
We now start to evaluate the commutator in (\ref{pot2end}). Note that using our current abbreviations $A$ and $B$ we obtain
\begin{equation} \nonumber
[\beta_2~\beta_1,u_{\rho}~u^{-1}]~=~-2~i~\left(%
\begin{array}{cc}
  0 & -B~\tau_3~\tau_1 \\
  B~\tau_3~\tau_1 & 0 \\
\end{array}%
\right),
\end{equation}
for the same reason as in (\ref{calcul}). Furthermore, our calculations yield the following results:
\begin{eqnarray}
\beta_2~[\beta_2~\beta_1,u_{\rho}~u^{-1}]&=&-2~i~\left(%
\begin{array}{cc}
  0 & \tau_2 \\
  \tau_2 & 0 \\
\end{array}%
\right)\left(%
\begin{array}{cc}
  0 & -B~\tau_3~\tau_1 \\
  B~\tau_3~\tau_1 & 0 \\
\end{array}%
\right)=\\&=&-2~i~B~\tau_2~\tau_3~\tau_1~\left(%
\begin{array}{cc}
I_2&0 \\
0&-I_2\\
\end{array}%
\right), \label{guliguli}
\end{eqnarray}
where the matrix product $\tau_2\tau_3\tau_1$ simplifies to
\begin{eqnarray}\nonumber
\tau_2~\tau_3~\tau_1&=& i~\left(%
\begin{array}{cc}
  0 & -\exp\left(-i~\varphi \right) \\
  \exp\left(i~\varphi \right) & 0
\end{array}%
\right)
\left(%
\begin{array}{cc}
  1 & 0 \\
  0 & -1
\end{array}%
\right)
\left(%
\begin{array}{cc}
  0 & \exp\left(-i~\varphi \right) \\
  \exp\left(i~\phi \right) & 0
\end{array}%
\right) \nonumber \\[1ex]
&=&i~I_2.
\end{eqnarray}
We substitute this result into (\ref{guliguli}) and get the commutator's final form:
\begin{eqnarray}\label{Bbeta}
\beta_2~[\beta_2~\beta_1,u_{\rho}~u^{-1}]&=&2~B~\beta \nonumber \\
&=& \frac{2~k~\cosh(2~k~\rho)~\beta}{\displaystyle{{\frac{k}{\varepsilon+m}~\cosh^2(k~\rho)+\frac{\varepsilon+m}{k}~\sinh^2(k~\rho)}}}
\nonumber \\
&=& \frac{2~k^2~\cosh(2~k~\rho)~\beta}{-m+\varepsilon~\cosh(2~k~\rho)}. \nonumber
\end{eqnarray}
On combining this with the definition (\ref{beta1}) and substituting into (\ref{pot3end}), we obtain
\begin{eqnarray}
V_1 &=& \left( \begin{array}{ccccc}
0 & 0 & 0 & {\displaystyle{\frac{1}{\rho}}}~\exp(-i~\varphi) \\
0 & 0 & {\displaystyle{\frac{1}{\rho}}}~\exp(i~\varphi) & 0 \\
0 & {\displaystyle{\frac{1}{\rho}}}~\exp(-i~\varphi) & 0 & 0 \\
{\displaystyle{\frac{1}{\rho}}}~\exp(i~\varphi) & 0 & 0 & 0
\end{array}
\right)+ \nonumber \\[2ex]
&+& \left( \begin{array}{ccccc}
\frac{2~k^2~\cosh(2~k~\rho)}{-m+\varepsilon~\cosh(2~k~\rho)}
 & 0 & 0 & 0 \\
0 & \frac{2~k^2~\cosh(2~k~\rho)}{-m+\varepsilon~\cosh(2~k~\rho)}
 & 0 & 0 \\
0 & 0 & -\frac{2~k^2~\cosh(2~k~\rho)}{-m+\varepsilon~\cosh(2~k~\rho)} & 0 \\
0 & 0 & 0 &-\frac{2~k^2~\cosh(2~k~\rho)}{-m+\varepsilon~\cosh(2~k~\rho)}
\end{array}
\right) \nonumber \\[2ex]
&=& \left( \begin{array}{ccccc}
\frac{2~k^2~\cosh(2~k~\rho)}{-m+\varepsilon~\cosh(2~k~\rho)} & 0 & 0 & {\displaystyle{\frac{1}{\rho}}}~\exp(-i~\varphi) \\
0 & \frac{2~k^2~\cosh(2~k~\rho)}{-m+\varepsilon~\cosh(2~k~\rho)} & {\displaystyle{\frac{1}{\rho}}}~\exp(i~\varphi) & 0 \\
0 & {\displaystyle{\frac{1}{\rho}}}~\exp(-i~\varphi) & -\frac{2~k^2~\cosh(2~k~\rho)}{-m+\varepsilon~\cosh(2~k~\rho)}& 0 \\
{\displaystyle{\frac{1}{\rho}}}~\exp(i~\varphi) & 0 & 0 &
-\frac{2~k^2~\cosh(2~k~\rho)}{-m+\varepsilon~\cosh(2~k~\rho)}
\end{array}
\right). \nonumber \\ \label{endv11}
\end{eqnarray}
This is the final form of our Darboux-transformed potential $V_1$. As before, the corresponding solution $\phi$ of the Dirac equation
(\ref{canalx}) can be constructed from the Darboux transformation (\ref{darbouxc1}).

\subsection{Physical interpretation}
Let us now briefly discuss physical aspects of the transformed potentials that were obtained in the previous examples.
To this end, recall the three-dimensional, stationary Dirac equation, minimally coupled to an electromagnetic field:
\begin{eqnarray}
\Bigg(\alpha_1~\Big(i~\partial_x+A_1\Big)+\alpha_2~\Big(i~\partial_y+A_2\Big)+\alpha_3~\Big(i~\partial_z+A_3\Big)
-\beta~m+E-\Phi \Bigg)~\Psi &=& 0, \nonumber
\end{eqnarray}
where the electric charge was set to one and the definitions for $\alpha_j$, $j=1,2,3$, and $\beta$ can be found in (\ref{alphabeta}). Furthermore,
$\Phi$ and $(A_1,A_2,A_3)$ denote the scalar and the vector component of the electromagnetic potential,
respectively. We will now rewrite this Dirac equation in cylindrical coordinates, first observe the obvious fact
that the equation can be written in the form
\begin{eqnarray}
\Bigg(i~\Big(\alpha_1~\partial_x+\alpha_2~\partial_y+\alpha_3~\partial_z \Big)+
\alpha_1~A_1+\alpha_2~A_2+\alpha_3~A_3-\beta~m+E-\Phi \Bigg)~\Psi &=& 0. \nonumber
\end{eqnarray}
The differential operators translate into cylindrical coordinates as in (\ref{canal}), that is, we have after
employing (\ref{sol18}) the following form of the Dirac equation:
\begin{eqnarray}
\Bigg(\beta_1~\partial_\rho+\frac{\beta_2}{\rho}~\partial_\varphi+
\alpha_1~A_1+\alpha_2~A_2+\alpha_3~A_3-\beta~m+E-\Phi\Bigg)~\psi=0, \label{intercyl}
\end{eqnarray}
where the definitions of $\beta_1,\beta_2$ and $\beta_3$ can be found in (\ref{beta1})-(\ref{beta3}). Before we
see our two examples in detail, let us evaluate the term involving the vector potential components $A_1,A_2$ and
$A_3$:
\begin{eqnarray}
\alpha_1~A_1+\alpha_2~A_2+\alpha_3~A_3 &=& \left(%
\begin{array}{cccc}
  0 & 0 & A_3 & A_1-i~A_2 \\
  0 & 0 & A_1+i~A_2 & -A_3 \\
  A_3 & A_1-i~A_2 & 0 & 0 \\
  A_1+i~A_2 & -A_3 & 0 & 0 \\
\end{array}
\right). \label{genmat}
\end{eqnarray}
We will now see that the two transformed potentials constructed in the above examples can be written in the form
(\ref{genmat}), such that the corresponding vector potential $(\Phi,A_1,A_2,A_3)$ is determined.
\paragraph{First example.} We compare the matrix of the transformed potential $V_1$, as given in
(\ref{endv1}), with the matrix (\ref{genmat}). Using Euler's formula, we first compare the off-diagonal elements, which
leads to the following identification of the vector potential
\begin{eqnarray}
\Phi~=~0 \qquad A_{1}~=~\displaystyle\frac{\cos{\phi}}{\rho}
\qquad A_{2}~=~\displaystyle\frac{\sin{\phi}}{\rho} \qquad A_{3}~=~0. \label{intera}
\end{eqnarray}
It remains to accomodate the diagonal elements of our matrix (\ref{endv1}). This can be done by observing that
the mass in the Dirac equation (\ref{intercyl}) is a coefficient of the matrix $\beta$ and can thus be joined with the
term (\ref{2bbeta}) of the transformed potential $V_1$. Thus, we introduce a position-dependent mass
$\hat{m}=\hat{m}(\rho)$ as follows:
\begin{eqnarray}
\hat{m} &=& m-2~B \nonumber \\
&=& m- \frac{4~m~(1+\log(\rho))}{4~m^2~\rho^2~\log^2(\rho)+1}. \label{intera2}
\end{eqnarray}
Hence, the transformed equation (\ref{canalx}) with $V_1$ given in (\ref{endv1}) corresponds to a Dirac equation with
position-dependent mass (\ref{intera2}), coupled minimally to a vector potential, the components of which are
given in (\ref{intera}).
\paragraph{Second example.} We proceed exactly as we did in the case of our first example. On comparing the matrices
(\ref{endv11}) and (\ref{genmat}), we obtain the same identifications (\ref{intera}) as before. The position-dependent mass
$\hat{m}=\hat{m}(\rho)$ reads this time
\begin{eqnarray}
\hat{m} &=& m-2~B \nonumber \\
&=& m- \frac{2~k^2~\cosh(2~k~\rho)}{-m+\varepsilon~\cosh(2~k~\rho)}. \label{intera3}
\end{eqnarray}
Hence, the transformed equation (\ref{canalx}) with $V_1$ given in (\ref{endv11}) corresponds to a Dirac equation with
position-dependent mass (\ref{intera3}), coupled minimally to a vector potential, the components of which are
given in (\ref{intera}).

Note that in general case \ref{aux} is equivalent the following
equation
\begin{equation}\label{matrix}
\left(M~\partial_x+N~\partial_y+E-V_0\right)~u=N~uC(y), \mbox{
where } E \mbox{ is arbitrary,}
\end{equation}
we can construct the exactly solvable transformed equation with
any energy value $E.$\\
But if $E$ is $4\times 4$ diagonal matrix $E=\Lambda=\left(%
\begin{array}{cccc}
  \lambda_1 & 0 & 0 & 0 \\
  0 & \lambda_2 & 0 & 0 \\
  0 & 0 & \lambda_3 & 0 \\
  0 & 0 & 0 & \lambda_4
\end{array}%
\right)$ we can construct the exactly solvable transformed equation
($V_0$ $\rightarrow$ $V_1$) with matrix solutions
$v=(\varphi_1,\varphi_2,\varphi_3,\varphi_4),$ where $\varphi_i$ is
solution to spinor transformed equation with $E=\lambda_i.$

Such as the initial equation with variable $E,$ the solutions of
transformed equation will with 'history'. We can consider appointed
initial energy and appointed final energy. Also it is possible to
construct solutions to transformed Dirac equation without 'history':
$v=Lv^0,$ $v^0$ is matrix solution to initial Dirac equation.

 We hope that such
results can be useful for study of electron transition, photon
absorption and creation.

\section{Conclusion}
The Darboux transformations for the general two-dimensional Dirac
equation are constructed. In fact, the Dirac equation in cylindrical
coordinates is particular case of the general two-dimensional Dirac
equation.  Our calculations allow to construct the exactly solvable three-dimensional
 Dirac equation for a potential that depends on two variables.
 Exactly solvable potentials of the
 Dirac equation in cylindrical coordinates are generated  by the Darboux
 transformation method, and the physical meaning of the transformed potentials is indicated.
\\\\
 This work is supported in part by contract of Russian Federal
Agency for Science and Innovations 02.740.11.5057.
  
\end{document}